\documentclass[preprint,showpacs,preprintnumbers,amsmath,amssymb]{revtex4}

\usepackage{epsfig}
\usepackage{graphicx}
\usepackage{dcolumn}
\usepackage{bm}

\def\DESepsf(#1 width #2){\epsfxsize=#2 \epsfbox{#1}}

\begin{document}


\title{Three-body Baryonic $\overline B\to \Lambda\,\bar p\,\pi$ Decays
 and Such}

\author{Chun-Khiang Chua}
\author{Wei-Shu Hou}%
\affiliation{Department of Physics,
National Taiwan University,
Taipei, Taiwan 10764,
Republic of China
}

\date{\today}

\begin{abstract}
We study decay rates and spectra of
$\overline B\to\Lambda\,\bar p\,\pi$,
$\Sigma^0\,\bar p\,\pi$,
$\Sigma^-\,\bar n\,\pi$,
$\Xi^0\,\overline \Sigma {}^+\,\pi$,
$\Xi^-\,\overline \Sigma {}^0\,\pi$ and
$\Xi^-\,\overline \Lambda\,\pi$ modes under a factorization approach.
The baryon pairs are produced through vector, axial vector,
scalar and pseudoscalar operators.
Previous predictions, including ours, 
are an order of magnitude too small compared to experiment.
By incorporating QCD counting rules and 
studying the asymptotic behavior,
we find an earlier relation between the 
pseudoscalar and axial vector form factors to be too restrictive.
Instead, the pseudoscalar and scalar form factors are 
related asymptotically. 
Following this approach, the measured 
$\Lambda\,\bar p\,\pi^+$ rate ($\sim 4.0 \times 10^{-6}$)
and spectrum can be understood, 
and $\Lambda$ should be dominantly left-hand polarized,
while we expect
${\mathcal B}(\Sigma^0\,\bar p\,\pi^+)\simeq 1.6\times 10^{-6}$.
These results and other predictions can be checked soon.
\end{abstract}

\pacs{13.25.Hw,  
      14.40.Nd}  

\maketitle

\section{Introduction}

Several three-body baryonic $B$ decays such as 
$\bar B\to D^{*}\,p\,\bar n$~\cite{Anderson:2000tz}, 
$p\,\bar p\,K$~\cite{Abe:2002ds} and
$D^{*}\,p\,\bar p$~\cite{Abe:2002tw} have emerged recently, even
though there is only one single two-body baryonic mode 
$\bar B^0 \to \Lambda^+_c\,\bar p$ 
that is observed so far~\cite{Abe:2002er,Gabyshev:2002dt}.
It has been argued that three-body baryonic modes could be
enhanced over two-body~\cite{Hou:2000bz}, by reducing energy
release to the baryons via emitting a fast recoil meson. One
consequence is enhancement near baryon pair threshold in
three-body modes. In our study of $B^0\to D^{*-}\,p\,\bar
n$~\cite{Chua:2001vh}, {\it assuming factorization}, we obtained
$\sim$ 60\% of experimental rate from the vector current
contribution, and the decay spectrum exhibits such threshold
enhancement. The same threshold enhancement effect was predicted
for the charmless $\rho\,p\,\bar n$ mode~\cite{Chua:2001xn}, and,
interestingly, the newly observed first ever charmless baryonic
mode, $B\to p\,\bar p\,K$, showed similar
feature~\cite{Abe:2002ds}. The measured decay rate can be understood to some
extent~\cite{Cheng:2001tr} and the spectrum can be reproduced by
using the factorization approach and QCD counting rule
arguments~\cite{Chua:2002wn}. Other charmless modes such as
$p\,\bar p\,\pi$, $\Lambda\,\bar p\,\pi$, $\Sigma^0\,\bar p\,\pi$
have been studied under the factorization
assumption and ${\mathcal B}(\Lambda\,\bar p\,\pi^+)=(3$--$5)\times10^{-7}$
and ${\mathcal B}(\Sigma^0\,\bar p\,\pi^+)=(0.8$--$1.8)\times10^{-6}$
were predicted~\cite{Cheng:2001tr,Chua:2002wn}. 

Recently, Belle reported~\cite{lambdappi}
\begin{equation}
{\mathcal B}(\Lambda\,\bar p\,\pi^+)
=(3.97^{+1.00}_{-0.80}\pm0.56)\times10^{-6},
\label{eq:expt}
\end{equation}
and ${\mathcal B}(\Sigma^0\,\bar p\,\pi^+)<3.8\times10^{-6}$ at
the 90\% confidence level. While the $\Lambda\,\bar p\,\pi$ decay
spectrum exhibits threshold enhancement as expected, the measured
rate turns out to be an order of magnitude higher than
predicted~\cite{Cheng:2001tr,Chua:2002wn}.
Furthermore, previous predictions placed 
${\mathcal B}(\Sigma^0\,\bar p\,\pi^+)$ considerably above
${\mathcal B}(\Lambda\,\bar p\,\pi^+)$.
If the factorization approach is not to be abandoned,
where could things go wrong?

We had noted that the $\overline B \to \Lambda\,\bar p\,\pi^+$ mode 
is sensitive to how one treats the vacuum to $\Lambda\bar p$ 
pseudoscalar matrix element~\cite{Chua:2002wn} 
under factorization. 
The analogous situation for meson case is known to be enhanced.
In this work, we revisit these two modes, as well as some SU(3)
related modes such as $\Sigma^-\,\bar n\,\pi$, $\Xi^0\,\overline
\Sigma {}^+\,\pi$, $\Xi^-\,\overline \Sigma {}^0\,\pi$ and
$\Xi^-\,\overline \Lambda\,\pi$. With the help of QCD counting
rules and taking into account the asymptotic behavior of baryonic
form factors, we can now account for the observed $\Lambda\,\bar
p\,\pi$ rate and spectra, where $\Lambda \bar p$ production
is dominated by the pseudoscalar density.
After improving the situation on the $\Lambda\,\bar p\,\pi$ rate,
we study the $\Lambda$ polarization, which is known to be useful
for constructing $CP$- and $T$-violation observables~\cite{Hou:2000bz}. 
We are able to make some predictions as well.

Our formulation is given in the next section, 
followed by results and discussion.

\section{Formalism}\label{sec:form}

Under the factorization assumption, the three-body baryonic $B$
decay amplitude consists of two parts. 
For one, the baryon pair is {\it current-produced} 
in association with a $B$ to meson transition.
For the other, the $B$ makes a {\it transition} to a baryon pair
and the recoil meson is current-produced~\cite{Chua:2002wn}. 
The $B\to p\,\bar p\,K$ mode receives both contributions, 
but the $\Lambda\,\bar p\,\pi^+$ mode, 
and analogously its SU(3) related modes such as 
$\Sigma^0\,\bar p\,\pi^+$, $\Sigma^-\,\bar n\,\pi^+$, 
$\Xi^0\,\overline \Sigma {}^+\,\pi^+$,
$\Xi^-\,\overline \Sigma {}^0\,\pi^+$ and 
$\Xi^-\,\overline \Lambda\,\pi^+$, 
receive {\it only} the current-produced contribution. 
We shall apply the term ``current-produced" to
scalar and pseudoscalar densities as well.

Take, for example, the $\overline B {}^0\to\Lambda\,\bar p\,\pi^+$
decay. Under factorization, the amplitude is~\cite{Chua:2002wn}
\begin{eqnarray}
{\cal M}(\Lambda\,\bar p\,\pi^+)&=&
 \frac{G_F}{\sqrt2}
   \langle \pi^+|\bar u\gamma^\mu (1-\gamma_5)b|\overline B{}^0 \rangle
   \bigg\{ (V_{ub} V_{us}^*a_1-V_{tb} V_{ts}^* a_4)
   \langle \Lambda\bar p\,|\bar s\gamma_\mu(1-\gamma_5) u|0\rangle
\nonumber\\
&& +2a_6 V_{tb}V_{ts}^*\,\frac{(p_{\Lambda}+p_{\bar p})_\mu}{m_b-m_u}
 \langle\Lambda\bar{p}\,| {\bar s (1 + \gamma_5) u}|0\rangle\bigg\}.
 \label{eq:Lambdappi}
\end{eqnarray}
The baryon pair $\Lambda\,\bar p$ is produced from 
vacuum through $\bar s\,\gamma_\mu\,(\gamma_5)u$ and 
$\bar s\,(\gamma_5)u$ operators, while 
$\overline B {}^0$ to $\pi^+$ transition is 
induced by $\bar u\gamma^\mu b$ current.
Note that, from isospin symmetry, we have $\langle \pi^0|\bar
u\gamma^\mu b|B^- \rangle =\langle \pi^+|\bar u\gamma^\mu
b|\overline B{}^0 \rangle/\sqrt2$, hence ${\cal M}(\Lambda\,\bar
p\,\pi^0)=
 {\cal M}(\Lambda\,\bar p\,\pi^+)/\sqrt2$.
For these current-produced modes, we
have
\begin{equation}
{\mathcal B}(B^-\to{\mathbf B}\overline{\mathbf B}^\prime\pi^0)
=\frac{\tau_{B^-}}{2\,\tau_{\overline B {}^0}}\,
{\mathcal B}(\overline B {}^0\to{\mathbf B}\overline{\mathbf B}^\prime\pi^+),
\label{eq:BBpi0}
\end{equation}
where $\tau_{\overline B {}^0,\,B^-}$ are the
$\overline B {}^0$ and $B^-$ meson lifetimes, 
and ${\mathbf B}$ stands for some baryon.

The (axial) vector current-produced matrix elements are decomposed as
\begin{eqnarray}
\langle {\rm\bf B}
\overline
{\rm \bf B}^\prime|V_{\mu}|0\rangle
&=&\bar{u}(p_{\rm\bf B}) \left\{(F_1+F_2)\gamma_{\mu}+
   \frac{\,F_2(t)}{m_{\rm\bf B}+m_{{\rm\bf B}^\prime}}
   \left(p_{\overline{\rm\bf B}^\prime}-p_{\rm\bf B}\right)_\mu\right\}
   v(p_{\overline{\rm\bf B}^\prime})\,,
\label{eq:vector_FF}\\
\langle {\rm\bf B}\overline{\rm\bf B}^\prime|A_{\mu}|0\rangle
&=&\bar{u}(p_{\rm\bf B})\left\{g_A\left(t\right)\gamma_{\mu}
   +\frac{h_A\left(t\right)}{m_{\rm\bf B}+m_{{\rm\bf B}^\prime}} \,
   \left(p_{\rm\bf B}+p_{\overline{\rm\bf B}^\prime}\right)_{\mu}\right\}
   \gamma_5\,
   v(p_{\overline{\rm\bf B}^\prime})\,,
\label{eq:axial_FF}
\end{eqnarray}
where $F_{1,2}$, $g_A$ and $h_A$ are the induced vector (Dirac and
Pauli), axial and the induced pseudoscalar form factors,
respectively,
and
$
 t\equiv (p_{\rm\bf B}+p_{\overline{\rm\bf B}^\prime})^2
 \equiv m^2_{{\mathbf B}\overline{\mathbf B}^\prime}
$.
The scalar and pseudoscalar matrix elements associated with the
$a_6$ term of Eq.~(\ref{eq:Lambdappi}) are expressed as
\begin{eqnarray}
\langle {\rm\bf B}\overline{\rm\bf B}^\prime|S|0\rangle
&=&f_S\left(t\right)\,
   \bar{u}(p_{\rm\bf B})
        v(p_{\overline{\rm\bf B}^\prime})\,,
\label{eq:scalar_FF}\\
\langle {\rm\bf B}\overline{\rm\bf B}^\prime|P|0\rangle
&=&g_P\left(t\right)\,
   \bar{u}(p_{\rm\bf B})
   \gamma_5\,
   v(p_{\overline{\rm\bf B}^\prime}).
\label{eq:psedoscalar_FF}
\end{eqnarray}
It is the $g_P(t)$ form factor that is the focus of
our attention, where we offer a refined discussion in face of
$B\to \Lambda\,\bar p\,\pi^+$ data.

The scalar and vector matrix elements can be related by the
equation of motion, $\langle {\rm\bf B}\overline{\rm\bf
B}^\prime|\partial^\mu V_{\mu}|0\rangle
 =i\,(m_q-m_{q^\prime})\langle {\mathbf B}\overline{\mathbf B}^\prime|
                     \bar q\,q^\prime|0\rangle$,
giving~\cite{Cheng:2001tr,Chua:2002wn}
\begin{equation}
f_S(t)=\frac{m_{\rm\bf B}-m_{{\rm\bf
B}^\prime}}{m_q-m_{q^\prime}}\,F_1(t). \label{eq:EQMS}
\end{equation}
%
%
We note that it is safe in the chiral limit 
$m_q$, $m_{q^\prime}\to 0$, and for $m_q\to m_{q^\prime}$ as well. 
For example, for $\langle \Lambda\bar
p|\bar s u|0\rangle$ we have
$({m_{\Lambda}-m_{p}})/({m_s-m_u})\sim 1$. For the modes studied
here, the factor $({m_{\rm\bf B}-m_{{\rm\bf
B}^\prime}})/(m_s-m_u)$ varies by $30,\,40\%$,
which illustrates SU(3) breaking.

The pseudoscalar and axial current matrix elements can be
analogously related. Using $\langle {\rm\bf B}\overline{\rm\bf
B}^\prime|\partial^\mu A_{\mu}|0\rangle
 =(m_q+m_{q^\prime})\langle {\mathbf B}\overline{\mathbf B}^\prime|
                     \bar q\,i\,\gamma_5\,q^\prime|0\rangle$,
we have
\begin{equation}
g_A(t)
+\frac{t}{(m_{\rm\bf B}+m_{{\rm\bf B}^\prime})^2}\,h_A(t)
=\frac{m_q+m_{q^\prime}}{m_{\mathbf B}+m_{{\mathbf B}^\prime}}\,g_P(t).
\label{eq:EQMP}
\end{equation}
As $m_{q},\,m_{q^\prime}\to 0$, we get $h_A(t)\to
-g_A(t)\,(m_{\mathbf B}+m_{{\mathbf B}^\prime})^2/t$
~\cite{Cheng:2001tr}. Since the $m_q/m_{\mathbf B}$ ratio is small, 
one is close to the chiral limit, hence the dependence of
$h_A(t)$ on $g_P(t)$ is weak.
However, to ensure good chiral behavior, we previously followed
Ref.~\cite{Cheng:2001tr} and took~\cite{Chua:2002wn}
\begin{equation}
g_P(t)=-g_A(t)\, 
\frac{m^2_{\rm GB}(m_{\mathbf B}+m_{{\mathbf B}^\prime})}
{(m_q+m_{q^\prime})(t-m^2_{\rm GB})},
\label{eq:mGB}
\end{equation}
where $m_{\rm GB}$ is the corresponding Goldstone boson (e.g. kaon) mass.
That is, $g_P(t)$ is obtained by changing the $1/t$ term in 
the asymptotic form of $h_A(t)$ to $1/(t-m^2_{\rm GB})$ 
and make use of Eq.~(\ref{eq:EQMP})~\cite{Cheng:2001tr}.
Indeed, Eq.~(\ref{eq:mGB}) gave too small a rate for 
$B\to\Lambda\,\bar p\,\pi^+$~\cite{Cheng:2001tr,Chua:2002wn}.
Due to the small quark-baryon mass ratio in Eq.~(\ref{eq:EQMP}),
we note that $g_A$ and $h_A$ are insensitive to $g_P$.
Therefore in the previous approach we need a very precise information on 
both $g_A$ and $h_A$, which is unavailable so far, to pinpoint $g_P$.
In this work we choose a different strategy by studying $g_{A,P}$ directly
and obtaining $h_A$ through Eq.~(\ref{eq:EQMP}).

According to QCD counting rules \cite{Brodsky:1974vy}, both the
vector form factor $F_1$ and the axial form factor $g_A$,
supplemented by leading logs, behave as $1/t^2$ in the
$t\to\infty$ limit. This is because we need two hard gluons to
impart large momentum transfer. Similarly, considering the
bilinear structure of the $S$ and $P$ operators, the scalar form
factor $f_S$ and pseudoscalar form factor $g_P$ also behave as
$1/t^2$ in the asymptotic limit. 
However, due to the need for helicity flip, 
one needs an extra $1/t$ for $F_2$ and $h_A$, 
hence they behave as $1/t^3$. 
We see that Eq.~(\ref{eq:mGB}) gives $1/t^3$ rather than $1/t^2$ 
asymptotic behavior for $g_P$, which is symptomatic.
In the electromagnetic current case, the
asymptotic form has been confirmed by many experimental
measurements of the nucleon magnetic form factor
$G^{p,n}_M=F^{p,n}_1+F^{p,n}_2$, over a wide range of momentum
transfers in the space-like region. The asymptotic behavior for
$G^p_M$ also seems to hold in the time-like region, as reported by
the Fermilab E760 experiment~\cite{Armstrong93} for
$8.9$~GeV$^2<t<13$~GeV$^2$. Another Fermilab experiment, E835, has
recently reported~\cite{Ambrogiani:1999bh} $G^p_M$ for momentum
transfers up to $\sim 14.4$~GeV$^2$. An empirical fit of $
\left|G^p_M\right|=C t^{-2} [\ln({t}/{Q_0^2})]^{-2} $ is obtained,
which is in agreement with the QCD counting rule.

\begin{table}[t!]
\caption{\label{tab:formfactor}Relations of baryon form factors
$F_1+F_2$, $g_A$ and $g_P$ with the nucleon magnetic form factors
$G_M^{p,n}$, $D_{A,P}$ and $F_{A,P}$ via the $(\bar s u)_{V,A,P}$
operators. Replacing $G^{p,n}_M$ by $G^{p,n}_E$ in the second
column, one obtains $F_1+F_2\,t/(m_{\mathbf
B}+m_{\overline{\mathbf B}^\prime})^2$.}
\begin{ruledtabular}
\begin{tabular}{lcc}
${\mathbf B}\overline{\mathbf B}^\prime$
          & $F_1+F_2$
          &$g_{A,\,P}$
          \\
\hline $\Lambda\,\bar p$
          & $-\sqrt{\frac{3}{2}}\,G^p_M$
          & $-\frac{1}{\sqrt6}\left(D+3F\right)_{A,P}$
          \\
$\Sigma^0\,\bar p$
          & $\frac{-1}{\sqrt{2}}\left(G_M^p+2\,G_M^n\right)$
          & $ \frac{1}{\sqrt{2}}\,\left(D-F\right)_{A,P}$
          \\
$\Sigma^-\,\bar n$
          & $-(G^p_M+2\,G^n_M)$
          & $(D-F)_{A,P}$
          \\
$\Xi^0\,\overline \Sigma {}^+$
          & $G^p_M-G^n_M$
          & $(D+F)_{A,P}$
          \\
$\Xi^-\,\overline \Sigma {}^0$
          & $\frac{1}{\sqrt2}\,\left(G^p_M-G^n_M\right)$
          & $ \frac{1}{\sqrt{2}}\,\left(D+F\right)_{A,P}$
          \\
$\Xi^-\,\overline \Lambda$
          & $\sqrt{\frac{3}{2}}\,\left(G^p_M+G^n_M\right)$
          & $-\frac{1}{\sqrt6}\left(D-3F\right)_{A,P}$
          \\
\end{tabular}
\end{ruledtabular}
\end{table}

The current induced form factors $F_1$, $F_2$ for the modes
studied here can be related to the nucleon (Sachs) magnetic and
electric form factors $G_{M,E}$, as shown in
Table~\ref{tab:formfactor}, where we also give the SU(3)
decomposition of $g_A$ and $g_P$ in terms of the form factors
$D_{A,P}$ and $F_{A,P}$. The $F_1+F_2$ terms are in fact obtained
by using
\begin{equation}
D_V=-\frac{3}{2}\,G^n_M, \qquad F_V=G^p_M+\frac{1}{2}\,G^n_M,
\label{eq:DVFV}
\end{equation}
with SU(3) decompositions similar to that of $g_{A,P}$. 
We can decompose $f_S$ similarly into $D_S$ and $F_S$, 
with (compare Eq.~(\ref{eq:EQMS}))
\begin{equation}
D_S=\frac{m_{\rm\bf B}-m_{{\rm\bf B}^\prime}}{m_q-m_{q^\prime}}\,
    \left(-\frac{3}{2}\,F^n_1\right),
\qquad 
F_S=\frac{m_{\rm\bf B}-m_{{\rm\bf B}^\prime}}{m_q-m_{q^\prime}}\,
    \left(F^p_1+\frac{1}{2}\,F^n_1\right).
\label{eq:DSFS}
\end{equation}
From the factorization assumption and Table~\ref{tab:formfactor}, 
we expect
\begin{eqnarray}
&&{\mathcal B}(\Lambda\,\bar p\,\pi^+)\sim
 2\,{\mathcal B}(\Lambda\,\bar p\,\pi^0),
\nonumber\\
&&{\mathcal B}(\Sigma^-\,{\bar n} \,\pi^{+})\sim
 2\,{\mathcal B}(\Sigma^-\,{\bar n} \,\pi^{0})\sim
 2\,{\mathcal B}(\Sigma^0\,\bar p\,\pi^+)\sim
 4\,{\mathcal B}(\Sigma^0\,\bar p\,\pi^0),
\nonumber\\
&&{\mathcal B}(\Xi^0\,\overline \Sigma {}^+\,\pi^{+})\sim
 2\,{\mathcal B}(\Xi^0\,\overline \Sigma {}^+\,\pi^{0})\sim
 2\,{\mathcal B}(\Xi^-\,\overline \Sigma {}^0\,\pi^{+})\sim
 4\,{\mathcal B}(\Xi^-\,\overline \Sigma {}^0\,\pi^{0}).
\end{eqnarray}

There is considerable data on the nucleon magnetic form factors.
This allows us to make a fit~\cite{Chua:2001vh}:
\begin{eqnarray}
G_M^p(t)=\sum^5_{i=1}\frac{x_i}{t^{i+1}}
\left[\ln\left(\frac{t}{\Lambda_0^2}\right)\right]^{-\gamma},
G_M^n(t)=-\sum^2_{i=1}\frac{y_i}{t^{i+1}}
\left[\ln\left(\frac{t}{\Lambda_0^2}\right)\right]^{-\gamma},
\label{eq:GM}
\end{eqnarray}
where $\gamma=2.148$,
$x_1=420.96$~GeV$^4$, $x_2=-10485.50$~GeV$^6$,
$x_3=106390.97$~GeV$^8$, $x_4=-433916.61$~GeV$^{10}$,
$x_5=613780.15$~GeV$^{12}$,
$y_1=292.62$~GeV$^4$, $y_2=-735.73$~GeV$^6$, and
$\Lambda_0=0.3$~GeV. They satisfy QCD counting rules and describe
time-like electromagnetic data such as $e^+e^-\to N \overline N$
suitably well. The data is extracted by assuming $|G^p_E|=|G^p_M|$
and $|G^n_E|=0$ (which gives better fit compared to the
$|G^n_E|=|G^n_M|$ case~\cite{Antonelli:fv}). With the fit of
Eq.~(\ref{eq:GM}), time-like $G^{p\,(n)}_M$ is real and positive
(negative)~\cite{Mergell:1996bf,Baldini:1999qn}. 
It is interesting to note that the fit coefficients $x_i$s 
alternate in sign, and likewise for $y_i$s.
Just two terms suffice for the latter because 
the neutron magnetic form factor data is 
relatively sparse~\cite{Chua:2001vh}. 
According to perturbative QCD~\cite{Lepage:1979za}, 
asymptotically ($t\to \infty$) one expects $G^n_M/G^p_M=-2/3$. 
We find that the fitted parameters for $G_M^n$ 
with the $|G^n_E|=0$ assumption gives 
$G^n_M/G^p_M\rightarrow-y_1/x_1=-0.70$, 
which is within 5\% of the QCD expectation. 
Note that, by use of $G_M=F_1+F_2$ and asymptotically $F_2/F_1\to1/t\to0$,
we have $F^n_1/F^p_1\to G^n_M/G^p_M\to -2/3$ as well.

The $F_2$ term can be related to $(G_E-G_M)/[t/(m_{\bf B}+m_{\bf
B^\prime})^2-1]$. However, we do not have much data on time-like
nucleon $G_E$. Thus, we concentrate on the $F_1+F_2$ term in
Eq.~(\ref{eq:vector_FF}) as we
did in Refs.~\cite{Chua:2001vh,Chua:2002wn}. We also use $G_M$ in
place of $F_1$ in Eqs.~(\ref{eq:EQMS}) and (\ref{eq:DSFS}).
The effect of the $F_2$ (or equivalently $G_E-G_M$) contribution
can be estimated by using form factor models such as Vector Meson
Dominance (VMD), where both $G_E$ and $G_M$ are available.

The time-like form factors related to $D_{A}$, $F_{A}$ are not yet
measured, but, as pointed out in Ref.~\cite{Cheng:2001tr},
their asymptotic behavior at $t\to\infty$ are known~\cite{Brodsky:1980sx}
and useful.
Asymptotically, they can be described by two form factors,
depending on the reacting quark having parallel or anti-parallel
spin with respect to the baryon spin~\cite{Brodsky:1980sx}. 
By expressing these two form factors in terms of $G_M^{p,n}$ 
as $t\to\infty$, one has
\begin{equation}
D_{A} \to G_M^p+\frac{3}{2}\,G_M^n,
\qquad
F_{A} \to \frac{2}{3}\,G_M^p-\frac{1}{2}\,G_M^n.
\label{eq:asymptoticA}
\end{equation}
%
%
In similar fashion, in
the asymptotic region the $f_S$ and $g_P$ form factors for the
chirality flip operators $S$ and $P$ can be expressed by {\it just
one} form factor, with spin of the interacting quark parallel to
the baryon spin. Anti-parallel spin corresponds to an
octet-decuplet instead of an octet-octet baryon pair. Since $g_P$
(equivalently $D_P,\ F_P$) and $f_S$ are related to the {\it same}
form factor, by following the approach of Ref.~\cite{Brodsky:1980sx}, 
as shown in Appendix A, we have
\begin{equation}
g_P \to f_S, \qquad \frac{D_{P(S)}}{F_{P(S)}} \to \frac{3}{2},
\label{eq:asymptoticP}
\end{equation}
as $t\to\infty$. 
This is a non-trivial requirement and it is not obeyed by
Eq.~(\ref{eq:mGB}). 
We note that Eq. (\ref{eq:asymptoticP})
is obtained without the use of the equation of motion.
The requirement of $D_S/F_S \to 3/2$
is consistent with Eq.~(\ref{eq:DSFS}),
which follows from Eq.~(\ref{eq:EQMS})
by using $F^n_1/F^p_1\to G_M^n/G_M^p\to-2/3$
asymptotically~\cite{Lepage:1979za}. Thus,
the use of the equation of motion for $f_S$ in Eq.~(\ref{eq:EQMS}) 
is consistent with the asymptotic relations in
Eq.~(\ref{eq:asymptoticP}).

The asymptotic relations hold for large $t$, hence they imply
relations on the leading terms of the corresponding form factors.
In general, more terms would be needed. 
In analogy to the neutron magnetic form factor case, 
we express $D_{A,P}$, $F_{A,P}$ up to 
the second term~\cite{Chua:2002wn},
\begin{eqnarray}
D_{A}(t)& \equiv & \left(\frac{\tilde d_1}{t^2}
                    +\frac{\tilde d_2}{t^3}\right)
 \left[\ln\left(\frac{t}{\Lambda_0^2}\right)\right]^{-\gamma},
\nonumber\\
F_{A}(t) & \equiv & \left(\frac{\tilde f_1}{t^2}
                    +\frac{\tilde f_2}{t^3}\right)
 \left[\ln\left(\frac{t}{\Lambda_0^2}\right)\right]^{-\gamma},
\nonumber\\
D_{P}(t)& \equiv &
 \left(\frac{\bar d_1}{t^2}+\frac{\bar  d_2}{t^3}\right)
 \left[\ln\left(\frac{t}{\Lambda_0^2}\right)\right]^{-\gamma},
\nonumber\\
F_{P}(t) & \equiv &
\left(\frac{\bar f_1}{t^2}+\frac{\bar f_2}{t^3}\right)
 \left[\ln\left(\frac{t}{\Lambda_0^2}\right)\right]^{-\gamma}.
\label{eq:FF}
\end{eqnarray}
The asymptotic relations of Eqs.~(\ref{eq:asymptoticA}), 
(\ref{eq:asymptoticP}) imply
$\tilde d_1 = x_1-3\,y_1/2$, $\tilde f_1 = 2\,x_1/3+y_1/2$,
$\bar d_1=(3y_1/2)[(m_{\rm\bf B}-m_{{\rm\bf B}^\prime})
 /(m_q-m_{q^\prime})]$
and
$\bar f_1=(x_1-y_1/2)[(m_{\rm\bf B}-m_{{\rm\bf B}^\prime})
 /(m_q-m_{q^\prime})]$,
%
while further information is needed to determine 
$\tilde d_2$, $\tilde f_2$, $\bar d_2$ and $\bar d_2$,
as we will discuss in the next section.
We note that the anomalous dimensions of $g_P$ and $f_S$ may not
be the same as that of $F_{1,2}$ and $g_A$. 
However, their effect is logarithmic hence not very important, 
and we apply the anomalous dimension of $F_1$ to others
for simplicity.

It is useful to compare with Refs.~\cite{Chua:2002wn,Cheng:2001tr}
on the treatment of $g_P(t)$ (or equivalently on $h_A(t)$), namely
Eq. (\ref{eq:mGB}). As a working assumption, this form of $g_P(t)$
with $m_{\rm GB}^2/(m_q+m_{q^\prime})$ factor was useful in
particular for the good behavior of the pseudoscalar matrix
element in the chiral limit. However, it may be too restrictive in
three aspects: $g_P(t)\propto g_A(t)$, which is too strong an
assumption; the appearance of the Goldstone boson pole in
time-like form factors, although $m_{\rm GB}$ is way below baryon
pair threshold; and a $1/t^3$ asymptotic behavior, rather than the
$1/t^2$ form as expected from QCD counting rules. Ultimately, it
does not satisfy the asymptotic relation of
Eq.~(\ref{eq:asymptoticP}). We have improved on these points in
our present treatment of $g_P(t)$.
%

\section{Results}

It is straightforward to use Eq.~(\ref{eq:Lambdappi})
to calculate $B\to \Lambda\,\bar p\,\pi$ and similar rates.
Before we start, let us first specify the parameters used.
We take $\phi_3$ (or $\gamma$) $=60^\circ$~\cite{Ciuchini:2000de}
and central values of $\vert V_{cb}\vert$
and $\vert V_{ub}\vert$ from Ref.~\cite{PDG}.
We use $m_{u(d)}/m_s=0.029\,(0.053)$, $m_s=120$~MeV and $m_b=4.88$~GeV
at $\mu=2.5$ GeV~\cite{PDG,Leutwyler:1996qg}.
The $B\to\pi$ transition form factor is 
given in Ref.~\cite{Melikhov:2000yu}.
For effective Wilson coefficients, we use $a_1=1.05$, 
$a_4\times 10^4=-387.3-121 i$ and $a_6\times 10^4=-555.3-121 i$
from Ref.~\cite{Cheng:1999xj} with $N_c=3$.

Following Ref.~\cite{Chua:2002wn},
we use the axial vector contribution to 
$B^0\rightarrow D^{*-}\,p\,\bar{n}$ decay
to constrain $\tilde f_2$ and $\tilde d_2$.
Since there is no scalar and pseudoscalar contribution
in this tree dominated mode, we simply use the chiral limit form
of $h_A(t)=-g_A(t)\,(m_p+m_n)^2/t$. 
The $g_P$ contribution is suppressed by the quark-baryon mass ratio.
We update our previous calculation~\cite{Chua:2001vh}
using the present input parameters,
finding the vector part of the branching ratio to be
${\mathcal B}_V(D^{*-}\,p\,\bar{n})
 =11.9\,(a^{\rm eff}_1/0.85)^2\times 10^{-4}$, where 
the same $a^{\rm eff}_1$ value as in Ref.~\cite{Cheng:2002fp} is used.
To reach the central value of the measured rate
${\cal B}(B^0\to D^{*-}\,p\,\bar{n}) =
(14.5^{+3.4}_{-3.0}\pm2.7)\times 10^{-4}$~\cite{Anderson:2000tz},
using $\tilde d_2+\tilde f_2=-956\,{\rm GeV}^6$~\cite{correction},
we find 
${\cal B}_A(D^{*-}\,p\,\bar{n})
 =2.6\,(a^{\rm eff}_1/0.85)^2\times 10^{-4}$ from the axial current.
Although the value of $\tilde d_2+\tilde f_2$ is about half of 
what was used in Refs.~\cite{Chua:2002wn} and \cite{Cheng:2002fp},
the change only affects the branching ratios of 
the charmless modes studied here at the $10^{-8}$ level.
Following Ref.~\cite{Chua:2002wn}, we use $\tilde d_2=\tilde f_2$.

With the axial contribution fixed, 
and with the scalar and vector contribution 
related by the equation of motion (Eq.~(\ref{eq:DSFS}))
we give in Table~\ref{tab:BR},
the vector plus scalar contribution (${\mathcal B}_V$) and 
the axial plus pseudoscalar contribution (${\mathcal B}_A$) to
$\overline B {}^0\to{\mathbf B}\overline {\mathbf B}^\prime\pi^+$
branching ratios.
For ${\mathcal B}_A$, we show two cases with either
vanishing or non-vanishing $\bar d_2$ and $\bar f_2$
from the pseudoscalar form factor, which is yet to be fixed.
Since the contribution from the vector plus scalar part 
does not interfere with the axial plus pseudoscalar part,
the branching fraction is a simple sum of the two, i.e.
${\mathcal B}={\mathcal B}_V+{\mathcal B}_A$,
just as for $B^0\rightarrow D^{*-}\,p\,\bar{n}$.
By using the relation of Eq.~(\ref{eq:BBpi0}),
${\mathcal B}({\mathbf B}\overline {\mathbf B}^\prime\pi^0)$
can be read off from Table~\ref{tab:BR} by a simple 1/2 factor.

\begin{table}[t]
\caption{\label{tab:BR}
Branching fractions for
${\mathbf B}\overline {\mathbf B}^\prime\pi^+$ modes
arising from the vector and scalar parts~(${\mathcal B}_V$),
and from the axial and pseudoscalar parts~(${\mathcal B}_A$).
The latter are given for the two cases of 
using the asymptotic $g_P$ ($\bar d_2=\bar f_2=0$) or 
the fitted $g_P$ ($\bar d_2=\bar f_2= -952\,{\rm GeV}^6$) 
from the $\Lambda\,\bar p\,\pi$ rate.
The branching fraction is
a simple sum of the two, i.e.
${\mathcal B}={\mathcal B}_V+{\mathcal B}_A$.
Rates for ${\mathbf B}\overline{\mathbf B}^\prime\pi^0$ modes are about
one half of those shown.}
%
\begin{ruledtabular}
\begin{tabular}{lccc}
Modes
           & ${\mathcal B}_V(10^{-6})$
           & \multicolumn{2}{c}{${\mathcal B}_A(10^{-6})$}
           \\
{}         &
           & use asymptotic $g_P$
           & use fitted $g_P$
           \\
\hline
$\Lambda\,\bar p\,\pi^+$
           & $0.13$
           & $7.97$
           & $3.84$
           \\
$\Sigma^0\,\bar p\,\pi^+$
           & $0.88$
           & $0.70$
           & $0.70$
           \\
$\Sigma^-\,\bar n\,\pi^+$
           & $1.79$
           & $1.41$
           & $1.41$
           \\
$\Xi^0\,\overline{\Sigma^+}\,\pi^+$
           & $0.17$
           & $2.23$
           & $1.20$
           \\
$\Xi^-\,\overline{\Sigma^0}\,\pi^+$
           & $0.09$
           & $1.14$
           & $0.63$
           \\
$\Xi^-\,\overline\Lambda\,\pi^+$
           & $0.15$
           & $0.38$
           & $0.20$
           \\
\end{tabular}
\end{ruledtabular}
\end{table}

We find ${\mathcal B}_{V}(\Lambda[\Sigma^0]\,\bar p\,\pi^+)=
0.13\,[0.88]\times10^{-6}$.
We note that ${\mathcal B}_{V}(\Lambda\,\bar p\,\pi^+)$
is consistent with previous studies~\cite{Chua:2002wn,Cheng:2001tr},
while ${\mathcal B}_{V}(\Sigma^0\,\bar p\,\pi^+)$ becomes 
slightly larger because of the different input values of
the neutron magnetic form factor parameters ($y_i$s).
Clearly, ${\mathcal B}_V(\Lambda\,\bar p\,\pi^+)$ part is 
still an order of magnitude
below the measured~\cite{lambdappi} branching ratio of 
Eq. (\ref{eq:expt}).
Before invoking the pseudoscalar form factor of Eq.~(\ref{eq:FF}),
let us make sure that other modifications are 
insufficient for the order of magnitude difference.

Recall that in the vector and scalar sector, we concentrated on
$F_1+F_2$ contributions without including the $G_E-G_M$ effect
since $G_E$ data is unavailable.
As noted earlier, one can try to estimate the $G_E-G_M$ effect 
by using some form factor model where both $G_E$ and $G_M$ are given.
We use a VMD model, Ref.~\cite{Mergell:1996bf},
which was discussed in our previous work~\cite{Chua:2001vh}.
Since $F^{\Lambda\bar p}_1(t)$ and $F^{\Lambda \bar p}_2(t)$
can be expressed in terms of $G_M^p$ and $G^p_E$,
and since the VMD model describes $G_M^p$ data better than $G_M^n$
(time-like) data~\cite{Mergell:1996bf},
perhaps the $\Lambda\,\bar p\,\pi^+$ mode may be a better place
to estimate the $G_E-G_M$ effect.
By incorporating VMD with the previous section
(following similar approach of Ref.~\cite{Chua:2001vh}),
we obtain ${\mathcal B}_V(\Lambda\,\bar p\,\pi^+)=0.27\times 10^{-6}$.
Although we gain by a factor of two compared to Table~\ref{tab:BR},
the effect is still of order $10^{-7}$, and is 
insufficient to account for the measured $\Lambda\,\bar p\,\pi^+$ rate.
The effect of $G_E-G_M$ is not likely to fill the gap between
${\mathcal B}_V(\Lambda\,\bar p\,\pi^+)$ and the measured
${\mathcal B}(\Lambda\,\bar p\,\pi^+)$.

We thus need to turn to the axial and pseudoscalar contributions.
Let us start by using only the $\bar d_1$ and $\bar f_1$ terms of $g_P$
determined by the asymptotic relation of Eq.~(\ref{eq:asymptoticP}),
i.e. taking $\bar d_2 = \bar f_2 = 0$.
It is remarkable that, as given in the first case for 
${\mathcal B}_A$ in Table~\ref{tab:BR} (column three),
the $1/t^2$ terms of $D_P$ and $F_P$ alone give 
${\mathcal B}(\Lambda\,\bar p\,\pi^+)\sim 8\times 10^{-6}$,
or overshooting the experimental value by a factor of two!
This is striking compared with the previous calculation 
using the ansatz of Eq. (\ref{eq:mGB}),
which gave results an order of magnitude 
too small~\cite{Chua:2002wn,Cheng:2001tr}.
Now, we know that the sign of $x_i$s and $y_i$s alternate
hence $G_M$ gets reduced as higher power (in $1/t$) terms are included.
We expect similar effect for $g_P$ by allowing for
nonzero $\bar d_2$ and $\bar f_2$.
We determine these coefficients (the $1/t^3$ terms) 
by fitting to the central value of the 
measured $\Lambda\,\bar p\,\pi^+$ rate.
We obtain
$-(\bar d_{2}+3\,\bar f_{2})/\sqrt{6} =1554.6\,{\rm GeV}^6$, 
which is displayed as the second case for ${\mathcal B}_A$
in Table~\ref{tab:BR}.
By assuming $\bar d_2\sim\bar f_2$,
we have $\bar d_2\sim-952\ {\rm GeV}^6$,
which has sign opposite to $\bar d_1$,
and is about twice the size of
$\tilde d_2=\tilde f_2=-478\,{\rm GeV}^6$,
the $1/t^3$ coefficients for the axial vector form factor.

\begin{figure}[t]
\includegraphics[width=3.3in]{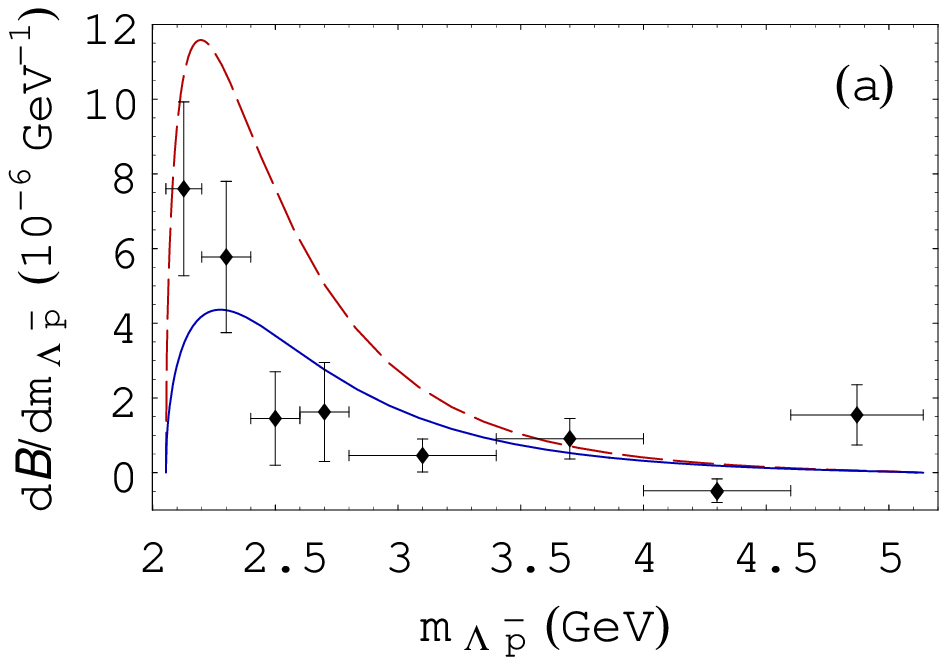}%
\includegraphics[width=3.3in]{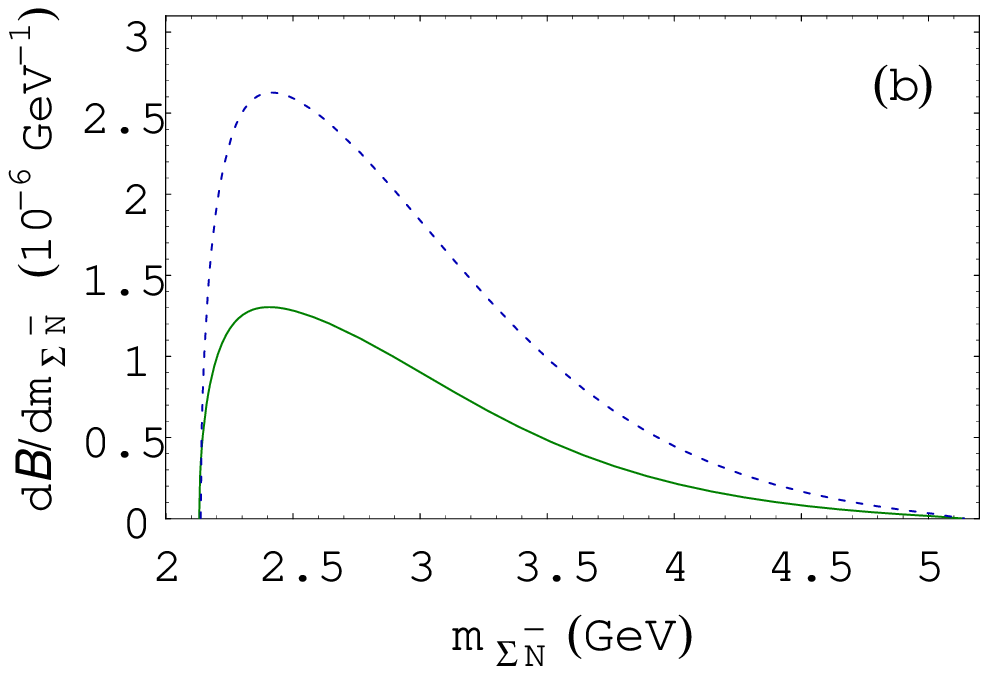}%
\caption{\label{fig:lambdappi}
(a) $d{\mathcal B}(\Lambda\,\bar p\,\pi^+)/dm_{\Lambda\bar p}$ spectrum,
where solid~(dashed) line is for using
the fitted (asymptotic) $g_P$ of $\bar d_2=\bar f_2=-952 \,{\rm GeV}^6\,(0)$;
(b) $d{\mathcal B}(\Sigma\,\bar N\,\pi^+)/dm_{\Sigma\bar N}$ spectra,
where solid (dotted) line is for $\Sigma^0\bar p$ ($\Sigma^-\bar n$).
The plots for $\pi^+$ replaced by $\pi^0$ are expected
to be similar but factor of 2 lower.}
\end{figure}

We show in Fig.~\ref{fig:lambdappi}(a) the $\Lambda\,\bar p\,\pi^+$
decay spectrum.
It is interesting that 
the predicted spectra in both $\bar d_2=\bar f_2=0$ and 
$\bar d_2=\bar f_2=-952\,{\rm GeV}^6$ cases are close to data.
The data suggests a curve between these two,
which conforms with our expectation that
the third, $1/t^4$ term would have same sign as $1/t^2$ term.
In future as the measured spectrum improved, one may in turn 
use it to extract baryon time-like from factors.

While ${\mathcal B}(\Lambda\,\bar p\,\pi^+)$ is enhanced from
the previous results~\cite{Chua:2002wn,Cheng:2001tr} 
by using our new approach to pseudoscalar $g_P$ form factor,
the enhancement in ${\mathcal B}(\Sigma^0\,\bar p\,\pi^+)$
turns out to be rather mild.
This can be understood from the relative weight of 
$\Lambda$ vs. $\Sigma^0$ in Eq. (\ref{eq:weight}) of Appendix A.
We expect ${\mathcal B}(\Sigma^0\,\bar p\,\pi^+)=1.6\times 10^{-6}$,
which is within the present Belle limit of
${\mathcal B}(\Sigma^0\,\bar p\,\pi^+)<3.8\times10^{-6}$
at 90\% confidence level~\cite{lambdappi}.
Furthermore, it is easy to verify the SU(3) predictions of
${\mathcal B}(\Sigma^-\,{\bar n} \,\pi^{+})\sim
 2\,{\mathcal B}(\Sigma^0\,\bar p\,\pi^{+})$
and
${\mathcal B}(\Xi^0\,\overline \Sigma {}^+\,\pi^{+})\sim
 2\,{\mathcal B}(\Xi^-\,\overline \Sigma {}^0\,\pi^{+})$
given in Table~\ref{tab:BR}.

\begin{figure}[t]
\includegraphics[width=3.3in]{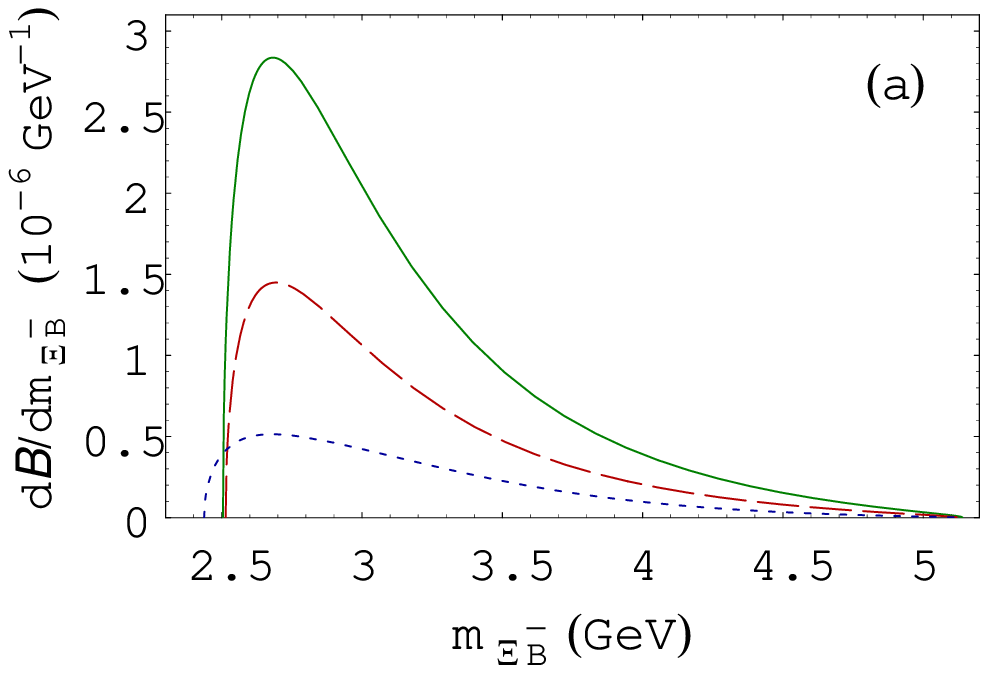}%
\includegraphics[width=3.3in]{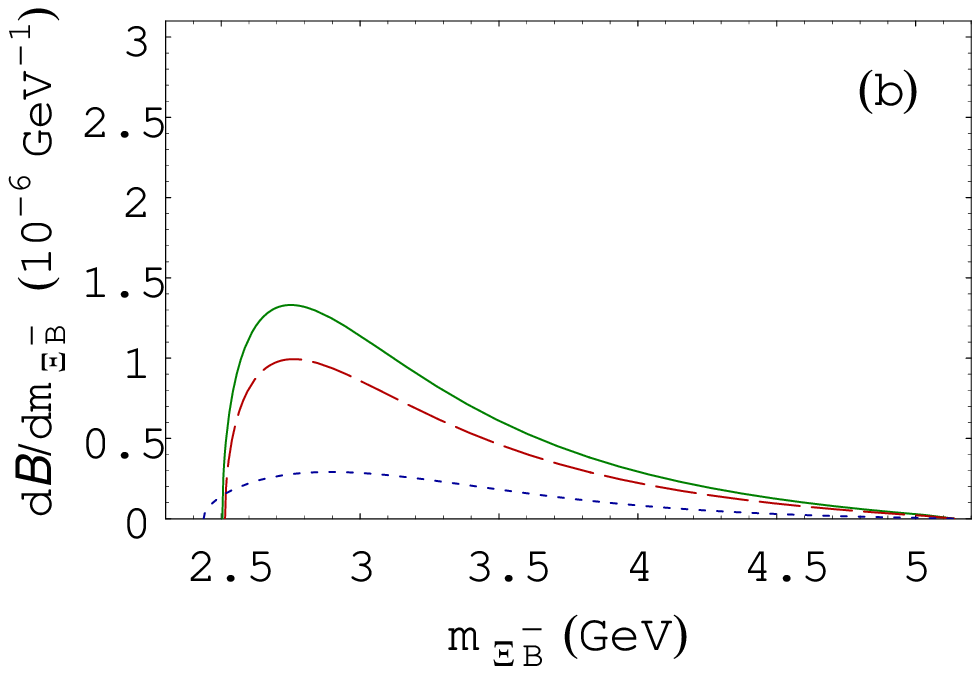}%
\caption{\label{fig:xibpi}
Solid, dashed and dotted lines are for
$d{\mathcal B}(\Xi^0\,\overline{\Sigma^+}\,\pi^+)
 /dm_{\Xi^0\,\overline{\Sigma^+}}$,
$d{\mathcal B}(\Xi^-\,\overline{\Sigma^0}\,\pi^+)
 /dm_{\Xi^-\,\overline{\Sigma^0}}$
and
$d{\mathcal B}(\Xi^-\,\overline{\Lambda}\,\pi^+)
 /dm_{\Xi^-\,\overline{\Lambda}}$,
respectively, 
for using
(a) the asymptotic $g_P$ ($\bar d_2=\bar f_2=0$), and
(b) the fitted $g_P$ ($\bar d_2=\bar f_2=-952 \,{\rm GeV}^6$).}
\end{figure}

In Fig.~\ref{fig:lambdappi}(b) we plot the 
$\Sigma^0\,\bar p\,\pi^+$ and $\Sigma^-\,\bar n\,\pi^+$ decay spectra.
The $\Sigma^0\,\bar p\,\pi^+$ spectrum is close to our previous
calculation in Ref.~\cite{Chua:2002wn}.
Since the corresponding SU(3) decomposition for 
these two mode is $D_P-F_P$, 
the rates are not sensitive to
$\bar d_2$ and $\bar f_2$ being zero or finite, 
so long that they are not too different from each other.
We show in Fig.~\ref{fig:xibpi} the
$\Xi^0\,\overline{\Sigma^+}\,\pi^+$,
$\Xi^-\,\overline{\Sigma^0}\,\pi^+$
and
$\Xi^-\,\overline{\Lambda}\,\pi^+$ decay spectra
with $\bar d_2$ and $\bar f_2$ zero or finite.

We expect Figs.~\ref{fig:lambdappi} and \ref{fig:xibpi} 
to give also the spectra of modes with $\pi^+$ replaced by $\pi^0$,
but with a factor of two reduction in rate from isospin factor.

In these three-body modes quite often we have a $\Lambda$ hyperon produced,
which is well known to self-analyse its spin upon decay 
and provides useful information for possible $CP$- and $T$-violation 
and chirality studies in $B$ decays~\cite{Hou:2000bz,Suzuki:2002cp}.
Following Ref.~\cite{Suzuki:2002cp}, the angular distribution of the cascade 
$\overline B\to\Lambda\,\bar p\,\pi\to \pi^-\, p\,\bar p\,\pi$ decay can be written as
\begin{equation}
\frac{d^2\Gamma}{d E_\Lambda d\cos\theta}=
\frac{1}{2}\frac{d\Gamma}{dE_\Lambda}
[1+\overline\alpha_\Lambda(E_\Lambda) \cos\theta],
\end{equation}
where $E_\Lambda$ is the $\Lambda$ energy measured in 
the $\overline B$ rest frame and $\theta$ is the supplementary angle 
between the emitted proton momentum and the $\overline B$ momentum 
in the $\Lambda$ rest frame.
We have $\overline\alpha_\Lambda(E_\Lambda)={\cal P}_\Lambda(E_\Lambda)\,\alpha_\Lambda$,  
where the $\Lambda$ polarization ${\cal P}_\Lambda(E_\Lambda)$ is given in Appendix B
and the $\alpha_\Lambda=0.642\pm0.013$~\cite{PDG} is the well-measured
$\Lambda$ decay asymmetry parameter.

\begin{figure}[t]
\includegraphics[width=3.3in]{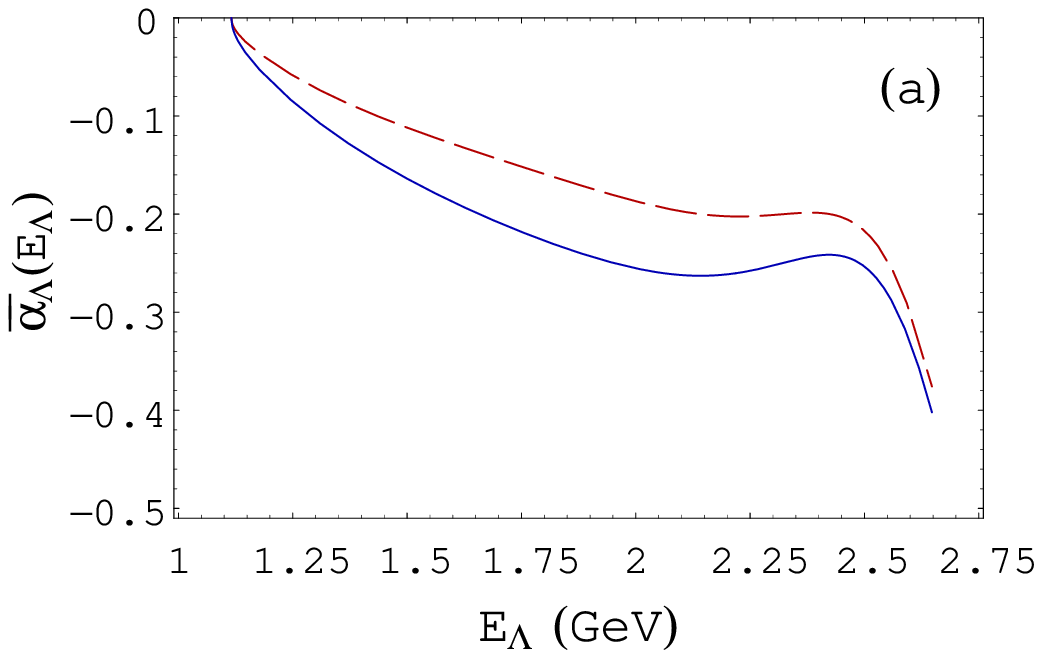}%
\includegraphics[width=3.3in]{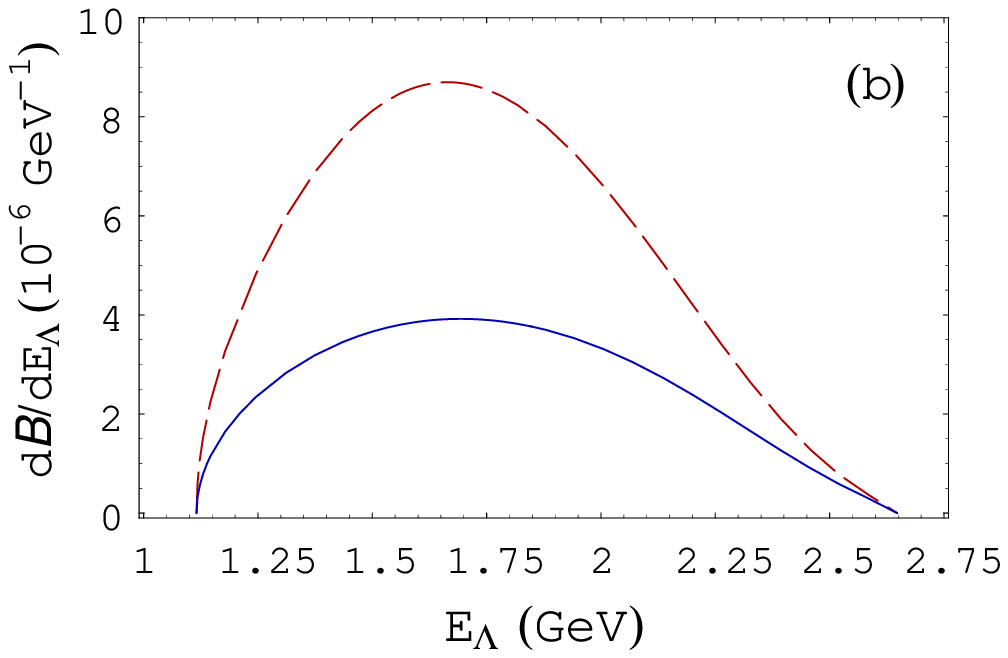}%
\caption{\label{fig:polarization}
(a) $\overline\alpha_\Lambda(E_\Lambda)$,
(b) $d{\mathcal B}(\Lambda\,\bar p\,\pi^+)/dE_\Lambda$ spectrum,
    where solid~(dashed) line is for using
the fitted (asymptotic) $g_P$ of $\bar d_2=\bar f_2=-952 \,{\rm GeV}^6\,(0)$.
}
\end{figure}

We show in Fig.~\ref{fig:polarization} the asymmetry
$\overline\alpha_\Lambda(E_\Lambda)$
and $d{\mathcal B}(\Lambda\,\bar p\,\pi^+)/dE_\Lambda$ spectrum.
The $\overline\alpha_\Lambda(E_\Lambda)$ plot is similar to the plot shown
in Ref.~\cite{Suzuki:2002cp} obtained by using some general arguments.
The negative $\overline\alpha_\Lambda(E_\Lambda)$ corresponds to
a left-handed helicity dominated $\Lambda$ in $B$ decay.
It is interesting to note that although the decay rate is dominated by the 
pseudoscalar term, we still have polarized $\Lambda$.
This can be understood by noting that the ratio of
scalar and pseudoscalar contributions is roughly given by the averaged
$f^2_S/g_P^2$, which is about 0.1, 
while the polarization ${\cal P}_\Lambda$ is roughly
given by the averaged $-2 f_S\, g_P/(f_S^2+g_P^2)\sim-2 f_S/g_P$ 
which can be as large as $-0.6$.  
The sharp turn of $\overline\alpha_\Lambda(E_\Lambda)$ towards 
much negative value for $E_\Lambda>2.5$~GeV is due to the fact that  
as $E_\Lambda$ increases, phase space
quickly reduces to a high $m_{\Lambda\bar p}$ region, 
resulting in the approach of $g_P$ to $f_S$ 
and consequently the increase in left-handed $\Lambda$ polarization.
It is well known that the $\Lambda$ spin is mainly carried by the
$s$ quark (as shown in Eq. (A2)) and it is left-handed 
in the $\overline B\to\Lambda\,\bar p\,\pi$ decay (as shown in Eq.~(\ref{eq:Lambdappi})).
Therefore, a dominantly left-handed $\Lambda$ reflects 
the $V-A$ nature of the weak interaction~\cite{Suzuki:2002cp}.
By comparing Fig.~\ref{fig:polarization}(a) and Fig.~\ref{fig:polarization}(b),
we find $-\overline\alpha_\Lambda\sim 0.2$--$0.3$ for 
the main portion of $\Lambda\,\bar p\,\pi$ events.
One should be able to check the sign of this asymmetry experimentally 
in the near future.

\section{Discussion and Conclusion}

Let us check the $\phi_3$ dependence of the modes considered here.
For all modes, ${\mathcal B}_V$ increases as 
we change $\phi_3$ from $60^\circ$ to $90^\circ$; on the other hand,
${\mathcal B}_A$ increases for
$\Lambda\,\bar p\,\pi$, $\Sigma\,\overline N\,\pi$
and $\Xi^-\,\overline \Lambda\,\pi$, but decreases for
$\Xi^0\,\overline \Sigma^+\,\pi$
and $\Xi^-\,\overline \Sigma^0\,\pi$.
However, the variations are at $10^{-7}$ order and far
less significant compared to the $K\pi$ case~\cite{He:1999mn}.
Since $a_6\,V_{tb}\,V_{ts}^*$ terms dominate,
we do not expect strong dependence on $\phi_3$ or $N_c$.
Similarly, single term dominance implies direct CP violation
cannot be large, which are found to be within $+5$\% for all modes.

It is interesting to discuss the implication on
$p\,\bar p\, K$ and $p\,\bar p\,\pi$
modes calculated in Ref.~\cite{Cheng:2001tr,Chua:2002wn}.
First of all, the changes are in the current-produced parts,
whereas these modes contain transition parts as well.
In particular, the $p\,\bar p\,\pi^-$ mode is dominated by transition.
From Eq.~(\ref{eq:EQMP}),
we see that $h_A(t)$ is close to its chiral limit form
because the dependence on $g_P$ is rather weak,
and $h_A(t)$ for the present work is similar to 
previous~\cite{Cheng:2001tr,Chua:2002wn}.
Therefore, the axial vector contributions to $p\,\bar p\,K$
and $p\,\bar p\,\pi$ modes are not affected.
The effect of $g_P$ only enter through the pseudoscalar term.
Since the pseudoscalar matrix element for $B\to p\,\bar p\,K$ decay,
$\langle p\bar p\,|(\bar s s)_P|0\rangle$~\cite{Chua:2002wn},
is Okubo-Zweig-Iizuka~(OZI) suppressed, 
we do not expect much change in these modes.
On the other hand, 
for $\langle p\bar p\,|(\bar d d)_P|0 \rangle$ 
of the $p\,\bar p\,\pi^-$ mode,
by using SU(3) and OZI argument as in Ref.~\cite{Chua:2002wn},
corresponds to $F_P-D_P$ and is non-negligible.
However, this mode is tree and transition dominant,
hence we still do not expect much change in rate~\cite{Chua:2002wn}.
Note that the transition form factor has a $1/t^3$ behavior.
For large enough $t$, the transition part is power suppressed.
We thus expect to see some $1/t^2$ contribution from the new $g_P$ term,
resulting in a slightly broader spectrum 
than previous~\cite{Chua:2002wn}.

In conclusion,
we study decay rates and spectra of
$\overline B \to \Lambda\,\bar p\,\pi$,
$\Sigma^0\,\bar p\,\pi$,
$\Sigma^-\,\bar n\,\pi$,
$\Xi^0\,\overline \Sigma {}^+\,\pi$,
$\Xi^-\,\overline \Sigma {}^0\,\pi$ and
$\Xi^-\,\overline \Lambda\,\pi$ modes, and the $\Lambda$ polarization in this work.
By suitably incorporating the asymptotic behavior of the baryonic
pseudoscalar matrix element,
we are able to obtain the $\Lambda\,\bar p\,\pi^+$ rate 
(in part by a fit) and spectrum close to experimental measurements.
The discrepancy between experimental~\cite{lambdappi} and
previous theoretical~\cite{Cheng:2001tr,Chua:2002wn} results is perhaps resolved.
While the $\Lambda\,{\bar p}\,\pi^+$ rate is enhanced from the
previous calculation,
we expect
${\mathcal B}(\Sigma^0\,\bar p\,\pi^+)=1.6\times 10^{-6}$,
which is within the present experimental limit and can be checked soon.
Although the $\Lambda\,\bar p\,\pi^+$ rate is 
dominated by the pseudoscalar term, we still have 
$\Lambda$ polarized giving $\overline\alpha_\Lambda\sim-(0.2$--$0.3)$.
The impact on $p\,\bar p\,K$ due to the present treatment of
the pseudoscalar form factor is negligible,
while we expect a slight broadening of the $p\,\bar p\,\pi^-$ spectrum.
Most of the subtleties in these modes come from 
the axial and especially the pseudoscalar form factors.
Information on these form factors may be 
obtained from studying these modes.
However, the underlying factorization assumption 
needs to be checked separately.
It is interesting that factorization seems to work in
$\overline B^0 \to D^{+}K^-K^0$ and $D^{*+}K^-K^0$ modes, where 
axial parts are absent and vector parts are known~\cite{Chua:2002pi}.
For these current-produced three-body baryonic modes,
we expect
${\mathcal B}({\mathbf B}\overline {\mathbf B}^\prime \pi^+)
 \sim
 2\, {\mathcal B}({\mathbf B}\overline {\mathbf B}^\prime \pi^0)$
as a consequence of factorization,
which does not depend on the complexity of baryonic form factors.

\begin{acknowledgments}
We thank H.-C. Huang, S.-Y. Tsai and M.-Z. Wang for discussions.
This work is supported in part by the National Science Council of
R.O.C. under Grants NSC-91-2112-M-002-027
and NSC-91-2811-M-002-043,
the MOE CosPA Project, and the BCP Topical Program of NCTS.
\end{acknowledgments}

\appendix
\section{Asymptotic relations for form factors $f_S$ and $g_P$}

We follow Ref.~\cite{Brodsky:1980sx} to obtain the asymptotic relations
for $f_S$ and $g_P$.
The wave function of an octet baryon can be expressed as 
\begin{equation}
|{\mathbf B}\,;\uparrow\rangle\sim
\frac{1}{\sqrt3}(|{\mathbf B}\,;\uparrow\downarrow\uparrow\rangle
                +|{\mathbf B}\,;\uparrow\uparrow\downarrow\rangle
                +|{\mathbf B}\,;\downarrow\uparrow\uparrow\rangle), 
\end{equation}
i.e. composed of 13-, 12- and 23-symmetric terms,
respectively.
For ${\mathbf B}=p,\,n,\,\Sigma^0,\,\Lambda$, we have
\begin{eqnarray}
|p\,;\uparrow\downarrow\uparrow\rangle&=&
\left[\frac{d(1)u(3)+u(1)d(3)}{\sqrt6} u(2)
 -\sqrt{\frac{2}{3}} u(1)d(2)u(3)\right]
|\uparrow\downarrow\uparrow\rangle,
\nonumber\\
|n\,;\uparrow\downarrow\uparrow\rangle&=&
(-|p\,;\uparrow\downarrow\uparrow\rangle
\,\,{\rm with}\,\,u \leftrightarrow d),
\nonumber\\
|\Sigma^0\,;\uparrow\downarrow\uparrow\rangle&=&
\bigg[-\frac{u(1)d(3)+d(1)u(3)}{\sqrt3}\,s(2)
      +\frac{u(2)d(3)+d(2)u(3)}{2\sqrt3}\,s(1)
\nonumber\\
      &&\,\,+\frac{u(1)d(2)+d(1)u(2)}{2\sqrt3}\,s(3)\bigg]
|\uparrow\downarrow\uparrow\rangle,
\nonumber\\
|\Lambda\,;\uparrow\downarrow\uparrow\rangle&=&
\bigg[\frac{d(2)u(3)-u(2)d(3)}{2}\,s(1)
      +\frac{u(1)d(2)-d(1)u(2)}{2}\,s(3)\bigg]
|\uparrow\downarrow\uparrow\rangle,
\end{eqnarray}
for the corresponding 
$|{\mathbf B}\,;\uparrow\downarrow\uparrow\rangle$ parts, 
while the 12- and 23-symmetric parts can be obtained by permutation.
To be consistent with the SU(3) decompositions of Table I, 
our $\Lambda$ state has an overall negative sign with 
respect to that of Ref.~\cite{Brodsky:1980sx}.

Following Ref.~\cite{Brodsky:1980sx}, we have
\begin{eqnarray}
\langle {\mathbf B}(p)| {\cal O}|{\mathbf B}^\prime(p^\prime)\rangle
&=&\bar u(p)\left[\frac{1+\gamma_5}{2}\,F^{+}(t)
                 +\frac{1-\gamma_5}{2}\,F^{-}(t)\right] u(p^\prime),
\nonumber\\
F^{\pm}(t)&=&e^{(\pm)}_{\parallel}
               ({\cal O}:{\mathbf B}^\prime\to{\mathbf B})
               \,F_{\parallel}(t),
\end{eqnarray}
in the large $t$ limit. Quark mass dependent terms 
behave like $m_q/\sqrt{|t|}$ and are neglected.
For simplicity, we illustrate with the space-like case.
Coefficients of $F_{\parallel}$ for the
${\mathcal O}=\bar q_L q^\prime_R,\,\bar q_L q^\prime_R$ cases
are given by
\begin{eqnarray}
e^{+}_{\parallel}(\bar q_L q^\prime_R:{\mathbf B}^\prime\to{\mathbf B})
&=&\langle {\mathbf B};\,\downarrow\downarrow\uparrow|
Q[q^\prime(1,\uparrow)\to q(1,\downarrow)]
|{\mathbf B}^\prime\,;\uparrow\downarrow\uparrow\rangle
\nonumber\\
&&+\langle {\mathbf B};\,\uparrow\downarrow\downarrow|
Q[q^\prime(3,\uparrow)\to q(3,\downarrow)]
|{\mathbf B}^\prime\,;\uparrow\downarrow\uparrow\rangle,
\nonumber\\
e^{-}_{\parallel}(\bar q_L q^\prime_R:{\mathbf B}^\prime\to{\mathbf B})&=&0,
\nonumber\\
e^{\pm}_{\parallel}(\bar q_R q^\prime_L:{\mathbf B}^\prime\to{\mathbf B})
&=& 
e^{\mp}_{\parallel}(\bar q_L q^\prime_R:{\mathbf B}^\prime\to{\mathbf B}),
\label{eq:e}
\end{eqnarray}
where $Q[q^\prime(1(3),\uparrow)\to q(1(3),\downarrow)]$ change the 
parallel spin $q^\prime(1(3))|\uparrow\rangle$ part of 
$|{\mathbf B}^\prime;\uparrow\downarrow\uparrow\rangle$
to a $q(1(3))|\downarrow\rangle$ part. 
It is easy to see that flipping the anti-parallel spin
$|\downarrow\rangle$ part of 
$|{\mathbf B}^\prime;\uparrow\downarrow\uparrow\rangle$
to $|\uparrow\rangle$ will give a decuplet instead of an octet state.
Thus, we need to consider the parallel spin case only.
By using the above equations, it is straightforward to obtain
\begin{eqnarray}
e^{+}_{\parallel}(\bar u_L d_R: n\to p)&=&-\frac{5}{3},
\nonumber\\
e^{+}_{\parallel}(\bar u_L s_R: \Lambda\to p)&=&\sqrt{\frac{3}{2}},
\nonumber\\
e^{+}_{\parallel}(\bar u_L s_R: \Sigma^0\to p)&=&-\frac{1}{3\sqrt2}.
\label{eq:weight}
\end{eqnarray}

By using $S,P=\bar q_L q^\prime_R\pm\bar q_R q^\prime_L$ 
and Eq.~(\ref{eq:e}), we have
$e^{\pm}_{\parallel}(\bar q q^\prime:
 {\mathbf B}^\prime\to{\mathbf B})
=e^{+}_{\parallel}(\bar q_L q^\prime_R:
 {\mathbf B}^\prime\to{\mathbf B})$ and
$e^{\pm}_{\parallel}(\bar q\gamma_5 q^\prime:
 {\mathbf B}^\prime\to{\mathbf B})
=\pm e^{+}_{\parallel}(\bar q_L q^\prime_R:
 {\mathbf B}^\prime\to{\mathbf B})$.
Hence 
\begin{equation}
f_S=g_P=
e^{+}_{\parallel}(\bar q_L q^\prime_R:
 {\mathbf B}^\prime\to{\mathbf B})
\,F_{\parallel},
\end{equation}
in the large $t$ limit.
In terms of $D_{S(P)}$ and $F_{S(P)}$, we have
$f_S(g_P)
=D_{S(P)}+F_{S(P)}$,
 $-(D_{S(P)}+3F_{S(P)})/\sqrt6$,
 $(D_{S(P)}-F_{S(P)})/\sqrt2$
for ${\mathbf B}^\prime{\mathbf B}=np$, $\Lambda p$ 
and $\Sigma^0 p$ cases, respectively.
Accordingly,
\begin{equation}
D_S=D_P=-F_{\parallel}\,,\qquad
F_S=F_P=-\frac{2}{3}\,F_{\parallel}\,,
\end{equation}
which implies Eq.~(\ref{eq:asymptoticP}).

\section{decay rate and polarization formula}

For a three-body 
$B\to h {\rm\bf B}\overline{\rm\bf B}^{\prime}$ decay, where $h$ is a 
pseudoscalar meson and
${\mathbf B}$, $\overline {\mathbf B}^\prime$ is a baryon anti-baryon pair,
in general the amplitude can be written as
\begin{eqnarray}
{\cal M}\left(B\to h{\rm\bf B}\overline{\rm\bf B}^{\prime}\right)
&=&
\frac{G_F}{\sqrt{2}}\, \biggl\{ 
{\cal A}\,\bar u(p_{{\mathbf B}})\,/\!\!\!p_h
v(p_{\overline{\mathbf B}^{\prime}})+ {\mathcal B}\,\bar
u(p_{{\mathbf B}})\,/\!\!\!p_h\gamma_5 
v(p_{\overline{\mathbf B}^{\prime}})
\nonumber\\
& &\qquad\qquad
+\,{\cal C}\,
\bar u(p_{{\mathbf B}})
v(p_{\overline{\mathbf B}^{\prime}})
+ {\cal D}\,
\bar u(p_{{\mathbf B}})\gamma_5 v(p_{\overline{\mathbf B}^{\prime}})
\biggr\} \,.\label{eq:general}
\end{eqnarray}
The decay rate is given by
\begin{equation}
d\Gamma=\frac{1}{\left(2\,\pi\right)^3}\,\frac{1}{32\,m_B^3}
\left(\Sigma_{\lambda_{1,2}}\,\left|{\cal M}\right|^2\right)dm_{12}^2\,dm_{23}^2\,,
\label{eq:DecayRate}
\end{equation}
where we assign the baryon $\mathbf B$ as particle~1, 
the anti-baryon $\overline {\mathbf B}^\prime$ as
particle~2 and the meson $h$ as particle~3,
and $\lambda_{1(2)}=\pm 1$ the helicity of the (anti-)baryon
$\mathbf B$ ($\overline{\mathbf B}^\prime$).

If the baryon $\mathbf B$ is in a definite helicity state,
its spin direction 
will remain the same in either the $B$ meson or its own rest frames.
For the baryon $\mathbf B$ with energy $E_1$ 
(measured in the $B$ meson rest frame)
the density matrix in the spin (or helicity) space is given by
\begin{equation}
\rho(E_1)=\frac{1}{2}\left[1+{\cal P}_{\mathbf B}(E_1)\,\hat p_1\cdot \sigma\right], 
\end{equation}
where $\hat p_1$ is the unit vector pointing opposite to the direction of 
the $B$ meson momentum in the $\mathbf B$ baryon rest frame and
\begin{equation}
{\cal P}_{\mathbf B}(E_1)=\frac{\int dm^2_{23}\,\Sigma_{\lambda_{1,2}} (-)^{\lambda_1}
   |{\cal M}|^2}
  {\int dm^2_{23}\,\Sigma_{\lambda_{1,2}} |{\cal M}|^2}.
\end{equation}
It is straightforward to obtain:
\begin{eqnarray}
\Sigma_{\lambda_{1,2}}\,(-)^{\lambda_1}\left|{\cal M}\right|^2&=& 
G_F^2\,4\biggl\{
{\rm Re}({\cal A}\,{\cal B}^*) m_1 
\bigl(2 s_1\cdot p_3\,p_2\cdot p_3-m_3^2\,s_1\cdot p_2\bigr)
\nonumber\\
&&+ {\rm Re}\left({\cal A}\,{\cal D}^*-{\cal B}\,{\cal C}^*\right)
m_1 m_2\,s_1\cdot p_3
\nonumber\\
&&+ {\rm Re}\left({\cal A}\,{\cal D}^*+{\cal B}\,{\cal C}^*\right)
\bigl(s_1\cdot p_3\,p_1\cdot p_2-s_1\cdot p_2\, p_1\cdot p_3\bigr)
\nonumber\\
&&-{\rm Re}\left({\cal C}\,{\cal D}^*\right) m_1\,s_1\cdot p_2\biggr\},
\label{eq:deltasquared}
\\
\Sigma_{\lambda_{1,2}}\,\left|{\cal M}\right|^2&=& G_F^2\,2\biggl\{\biggl[
\left|{\cal A}\right|^2\bigl(2 p_1\cdot p_3\,p_2\cdot p_3-m_3^2\,p_1\cdot p_2-
m_1 m_2 m_3^2\bigr)
\nonumber\\
&&+ 2\,{\rm Re}\left({\cal A}\,{\cal C}^*\right)
\bigl(m_1 p_2\cdot p_3-m_2\,p_1\cdot p_3\bigr)
+\bigl|{\cal C}\bigr|^2
\bigl(p_1\cdot p_2-m_1 m_2\bigr)\biggr]
\nonumber\\
&&+\biggl[{\cal A}\to{\cal B},\,{\cal C}\to {\cal D},\,m_2\to-m_2\biggl]\biggr\}\,,
\label{eq:squared}
\end{eqnarray}
where $s_1$ is the helicity vector of the baryon
$\mathbf B$ (spinor) with $\lambda_1=+1$.
It is easy to check that by neglecting $m_1$ we have $m_1 s_1\to p_1$ 
and we obtain ${\cal P}_{\mathbf B}(E_1)\to-1$ in the fully left-handed chiral case 
(${\cal A}\sim -{\cal B}$ and ${\cal C}\sim {\cal D}$) 
as expected from Eq.~(\ref{eq:general}).
In general, the polarization ${\cal P}_{\mathbf B}(E_1)$ can be easily evaluated 
in the $B$ meson rest frame by using 
\begin{equation}
s_1=\frac{1}{m_1 \sqrt{(p_B\cdot p_1)^2-m_1^2 m_B^2}}
                  (p_B\cdot p_1\, p_1-m_1^2\,p_B),
\end{equation}
where $p_B$ is the momentum of the $B$ meson,
and the standard technique of expressing $p_B\cdot p_i$, $p_i\cdot p_j$ in terms of 
$m^2_{ij}$.
Given these formulas, the task is now reduced to extract the 
${\cal A}$--${\cal D}$ terms for an amplitude of interest.



\end{document}